\documentclass[11pt]{article}
\newcommand{\Comment}[1]{{}}
\usepackage[textwidth = 430 pt, textheight = 630 pt]{geometry}
\usepackage{amssymb,euscript,amsmath,amsfonts}

\usepackage{color}
\definecolor{MyDarkBlue}{rgb}{0.15,0.15,0.45}
\usepackage[linktocpage=true]{hyperref}
\hypersetup{
colorlinks=true,
citecolor=MyDarkBlue,
linkcolor=MyDarkBlue,
urlcolor=MyDarkBlue,
pdfauthor={Neil Lambert and Constantinos Papageorgakis},
pdftitle={Relating U(N)xU(N) to SU(N)xSU(N) Chern-Simons Membrane theories},
pdfsubject={hep-th}
}

\usepackage[numbers,sort&compress]{natbib}
\usepackage{hypernat}

\newcommand\ignore[1]{}
\def\one{{\,\hbox{1\kern-.8mm l}}}

\newcommand{\SO}{\mathrm{SO}} 
\newcommand{\SU}{\mathrm{SU}} \newcommand{\U}{\mathrm{U}}
\newcommand{\su}{{\mathfrak{su}}}
\newcommand{\uf}{ {\mathfrak{u}}}

\newcommand{\Cset}{{\,\,{{{^{_{\pmb{\mid}}}}\kern-.45em{\mathrm C}}}}}

\newcommand{\eg}{{\it e.g.}}
\newcommand{\be}{\begin{equation}}
\newcommand{\ee}{\end{equation}}
\newcommand{\bea}{\begin{eqnarray}}
\newcommand{\eea}{\end{eqnarray}}

\newcommand{\bZ}{{\bar Z}}

\begin{document}

\rightline{}
   \vspace{1.8truecm}

\vspace{15pt}

\centerline{\LARGE \bf {\sc Relating $\U(N)\times \U(N)$ to $\SU(N)\times \SU(N)$}} \vspace{.2truecm}
\centerline{\LARGE \bf {\sc Chern-Simons Membrane theories}} \vspace{2truecm} \thispagestyle{empty}
\centerline{
    {\large {\bf {\sc Neil Lambert}}}\footnote{E-mail address: \href{mailto:neil.lambert@kcl.ac.uk}{\tt neil.lambert@kcl.ac.uk}} { and}
    {\large {\bf{\sc Constantinos Papageorgakis}}}\footnote{E-mail address:
                                 \href{mailto:costis.papageorgakis@kcl.ac.uk}{\tt costis.papageorgakis@kcl.ac.uk}}
                                                           }

\vspace{.8cm}
\centerline{{\it Department of Mathematics}}
\centerline{{\it King's College London}}
\centerline{{\it The Strand, WC2R 2LS}}
\centerline{{\it London, UK}}

\vspace{2.0truecm}

%%%%%%%%%%%%%%%%%
\thispagestyle{empty}

\centerline{\sc Abstract}

\vspace{0.4truecm}
\begin{center}
\begin{minipage}[c]{380pt}{
\noindent By integrating out the $\U(1)_B$ gauge field, we show that the $\U(n)\times \U(n)$ ABJM theory at level $k$ is equivalent to a $\mathbb Z_k$ identification of the $(\SU(n) \times \SU(n))/\mathbb Z_n$ Chern-Simons theory, but only when $n$ and $k$ are coprime. As a consequence, the $k=1$ ABJM model for two M2-branes in $\mathbb{R}^8$ can be identified with the ${\cal N}=8$ $(\SU(2) \times \SU(2))/\mathbb Z_2$ theory. We also conjecture that the $\U(2)\times \U(2)$ ABJM model at $k=2$ is equivalent to the $\mathcal N=8$ $\SU(2)\times\SU(2)$-theory.}
\end{minipage}
\end{center}

\vspace{.4truecm}

\noindent

\vspace{.5cm}

\setcounter{page}{0}

\newpage

\setcounter{footnote}{0}

\linespread{1.1}
\parskip 4pt

{}~
{}~

\subsection*{Introduction}

There has been considerable activity in the past two years leading to a new class of highly supersymmetric
three-dimensional conformal Chern-Simons theories which control the dynamics of multiple M2-branes in M-theory.
This work started with the papers \cite{Bagger:2006sk,Bagger:2007jr,Gustavsson:2007vu,Bagger:2007vi}, which
were the first to construct interacting theories with the correct symmetries; ${\cal N}=8$ supersymmetry and
$\SO(8)$ R-symmetry. These theories have no continuous coupling constant but they do admit a discrete coupling
$k$ that arises as the level of the Chern-Simons terms. However, this model is only capable of potentially
describing two M2-branes and its spacetime interpretation is unclear.  The generalisation to an arbitrary
number of $n$ M2-branes in a $\mathbb{R}^8/\mathbb{Z}_k$ orbifold was provided by the celebrated ABJM models
\cite{Aharony:2008ug} which are $\U(n)\times \U(n)$ Chern-Simons-matter theories with ${\cal   N}=6$
supersymmetry and $\SU(4)$ R-symmetry.

The main aim of this note is to elucidate the relation between the ${\cal N}=6$ $\U(n)\times \U(n)$ ABJM models  and  $(\SU(n)\times \SU(n))/\mathbb Z_n$ theories. As already noted in \cite{Aharony:2008ug}, the relative $\U(1)_B$ gauge field of the ABJM theories can be naturally integrated out. Since $\U(n) \simeq (\U(1)\times \SU(n))/\mathbb Z_n$, naively the effect of this is to reduce the $\U(n)\times \U(n)$ theory to a $\mathbb{Z}_k$ quotient of the $(\SU(n)\times \SU(n))/\mathbb Z_n$ theory. However we will see that there is a global obstruction to this $\mathbb{Z}_k$ identification unless $n$ and $k$ are coprime.

We will be particularly interested in the case with $n=2$, where the Lagrangian is precisely the original proposal of \cite{Gustavsson:2007vu,Bagger:2007jr} and has ${\cal N}=8$ supersymmetry and $\SO(8)$ R-symmetry. According to the above, the $\mathcal N=6$ ABJM $\U(2)\times \U(2)$ theory can be mapped to the ${\cal N}=8$, $(\SU(2) \times \SU(2))/\mathbb Z_2$ theory along with the $\mathbb{Z}_k$ identification on the fields when $k$ is odd.  For $k=1$ the identification is trivial and hence the $(\SU(2) \times \SU(2))/\mathbb Z_2$-theory at $k=1$ describes two M2-branes in flat space.

We also seek to clarify statements in \cite{Lambert:2008et,Distler:2008mk} which computed the moduli space of the $\mathcal N=8$ theory and argued that it corresponded to the IR limit of an $\SO(5)$ orbifold in type IIA, obtained by including one unit of discrete torsion for the background 3-form gauge field. In fact the discussion in \cite{Lambert:2008et,Distler:2008mk} is insufficient to distinguish between the orbifolds with and without torsion since they both have the same moduli space. Our discussion here shows that at $n=k=2$ the ABJM model cannot be reduced to a $\mathbb{Z}_2$ quotient of the $(\SU(2) \times \SU(2))/\mathbb Z_2$ theory.  However, the $\mathcal N=8$ $\SU(2)\times \SU(2)$ theory at $k=2$ does give the correct moduli space. This, along with the similarity between the two Lagrangians leads us to conjecture that the $\SU(2)\times \SU(2)$ theory obtained from the Lagrangian of \cite{Bagger:2007jr,Gustavsson:2007vu,Bagger:2007vi} has an M-theory interpretation at $k=2$ and is equivalent to the $\U(2)\times \U(2)$ Chern-Simons theory of \cite{Aharony:2008ug}, corresponding to the IR fixed point of a 2+1d $\mathrm O(4)$ orbifold theory. These results should make the connection between the theories of \cite{Bagger:2006sk,Bagger:2007jr,Gustavsson:2007vu,Bagger:2007vi,Bagger:2008se} and ABJM transparent and explain any aspects of M-theory physics captured by the former.

Note that the Chern-Simons-matter Lagrangians are entirely determined by the 3-algebra data which includes the Lie algebra of the gauge group. In the quantum theory one must also specify the full global gauge group. This choice manifests itself by allowing for different flux quantization conditions which in turn yield distinct quantum theories, with the same symmetry algebra. To account for this we will label the Lagrangian by is Lie algebra but the associated quantum theories will be labeled by the global gauge group.

\subsection*{ ${\cal N}=6$ Chern-Simons theories from 3-algebras}

Let us start by considering the general form of three-dimensional Lagrangians with scale symmetry and ${\cal
N}=6$ supersymmetry \cite{Bagger:2008se}:
\begin{eqnarray}
\label{niceaction}
\nonumber {\cal L} &=& -{\rm Tr}(D_\mu Z^A,
D^\mu \bZ_A) -
i{\rm Tr}(\bar\psi^A,\gamma^\mu D_\mu\psi_A) -V+{\cal L}_{CS}\\[2mm]
&& -i{\rm Tr}(\bar\psi^A, [\psi_{A},
Z^B;\bZ_{B}])+2i{\rm Tr}(\bar\psi^A,[\psi_{B},Z^B;\bZ_{A}])\\
\nonumber &&+\frac{i}{2}\varepsilon_{ABCD}{\rm Tr}(\bar\psi^A,[
Z^C,Z^D;\psi^B]) -\frac{i}{2}\varepsilon^{ABCD}{\rm Tr}(\bZ_D,[\bar
\psi_{A},\psi_B;\bZ_{C}])\ ,
\end{eqnarray}
where
\begin{eqnarray}
  V &=& \frac{2}{3}{\rm Tr}(\Upsilon^{CD}_B,\bar\Upsilon^B_{CD}) \\
\nonumber  \Upsilon^{CD}_B &=&
  [Z^C,Z^D;\bZ_B]-\frac{1}{2}\delta^C_B[Z^E,Z^D;\bZ_E]+\frac{1}{2}\delta^D_B[Z^E,Z^C;\bZ_E]  ,
\end{eqnarray}
and ${\cal L}_{CS}$ is a Chern-Simons term that we will describe in
detail below. The bracket $[\cdot,\cdot;\cdot]$ is antisymmetric in
the first two entries and defines the triple product of the
3-algebra where the scalars and fermions take values. Introducing a
basis $T^a$ for the 3-algebra, so that $Z^A = Z^A_aT^a$, $\psi_A =
\psi_{Aa}T^a$, allows us to use structure constants defined through
\begin{equation}
[T^a,T^b;T_c] = f^{ab}{}_{cd}T^d\ .
\end{equation}
Here we use notation where complex conjugation raises and lowers
both $A$ and $a$ indices (whereas in \cite{Bagger:2008se} a raised
$a$ index was given a bar).

The supersymmetry transformations are
\begin{eqnarray}\label{finalsusy}
\nonumber  \delta Z^A_d &=& i\bar\epsilon^{AB}\psi_{Bd} \\
\nonumber  \delta \psi_{Bd} &=& \gamma^\mu D_\mu Z^A_d\epsilon_{AB}
+
  f^{ab}{}_{cd}Z^C_a Z^A_b \bZ_{C}^c \epsilon_{AB}+f^{ab}{}_{cd} Z^C_a Z^D_{b} \bZ_{B}^c\epsilon_{CD} \\
  \delta \tilde A_\mu{}^c{}_d &=&
-i\bar\epsilon_{AB}\gamma_\mu Z^A_a\psi^{Bb} f^{ca}{}_{bd} + i\bar\epsilon^{AB}\gamma_\mu \bZ_{Ab}\psi_{B}^a
f^{cb}{}_{ad}\ ,
\end{eqnarray}
where the covariant derivative is $D_\mu Z^A_d = \partial_\mu Z^A_d - \tilde  A_\mu{}^c{}_dZ^A_c$ and similarly
for the other fields.

One recovers the general form of the ABJM and ABJ Lagrangians \cite{Aharony:2008ug,Aharony:2008gk} by taking the 3-algebra to be $n\times m$ complex matrices with
\begin{equation}\label{ABJM3algebra}
[Z^A,Z^B;\bar Z_C] = -\frac{2 \pi}{k} (Z^AZ_C^\dag Z^B -Z^BZ_C^\dag Z^A)
\end{equation}
and introducing a metric on the 3-algebra
\begin{equation}\label{metric}
  {\rm Tr}(T_a ,T^b) = {\rm tr}(T_a^\dag T^b)\ ,
\end{equation}
where on the right hand side ${\rm tr}$ is the ordinary matrix trace.

The gauge symmetry is generated by
\begin{equation}
\delta Z^A = \Lambda^b{}_c [Z^A,T_b;T^c] = M_LZ^A-Z^A M_R\ ,
\end{equation}
where $M_L = \frac{2 \pi}{k}\Lambda^b{}_c T_b (T^c)^\dag$,  $M_R = \frac{2 \pi}{k}\Lambda^b{}_c (T^c)^\dag T_b$
and $ (\Lambda^b{}_c)^* = - \Lambda^c{}_b$. Thus we see that $M_{L/R}^\dag = -M_{L/R}$ and hence they can be
viewed as generators of $\uf(n)\times \uf(m)$ with $Z^A$ and $\psi_A$ in the bi-fundamental representation.

As a result, the action of the gauge fields $\tilde A_{\mu}^a{}_b$ on $Z^A_a$ can be respectively rewritten in terms of left- and right-acting $\mathfrak{u}(n)$ and $\mathfrak{u}(m)$ gauge fields $\tilde A^{L/R}_\mu$
\begin{equation}
D_\mu Z^A = \partial_\mu Z^A - i\tilde A^L_\mu Z^A + iZ^A\tilde A^R_\mu
\end{equation}
and the term ${\cal L}_{CS}$ in (\ref{niceaction}) is then a level $(k,-k)$ Chern-Simons term for $\uf(n)\times
\uf(m)$
\begin{equation}\label{CSterm}
{\cal L}_{CS} = \frac{k}{4\pi}\varepsilon^{\mu\nu\lambda}\left({\rm tr}(\tilde A^L_\mu\partial_\nu \tilde
A^L_\lambda - \frac{2}{3}\tilde A^L_\mu \tilde A^L_\nu \tilde A^L_\lambda) -{\rm tr}(\tilde
A^R_\mu\partial_\nu\tilde  A^R_\lambda - \frac{2}{3}\tilde A^R_\mu \tilde A^R_\nu \tilde A^R_\lambda)\right)\ .
\end{equation}
The Chern-Simons level $k$ is integer whenever tr is the trace in
the fundamental representation.

However, it is important to note that ${\rm tr}(M_L)={\rm tr}(M_R)$. Thus if $M_L=i\theta_L\one_{n\times n}$
and $M_R=i\theta_R\one_{m\times m}$, we have $n\:\theta_L=m\:\theta_R$. Since the action of these Abelian
$\U(1)$'s is $Z^A\to e^{i\theta_L}Z^Ae^{-i\theta_R}=e^{i(\theta_L-\theta_R)}Z^A$, these cancel for the ABJM
case of $m=n$ and hence the gauge algebra is really $\su(n)\oplus \su(n)\oplus \uf(1)_Q$, where the overall
$\U(1)_Q$ acts trivially on all  fields. This is not true in the ABJ case, where $m\ne n$ and the gauge group
is an honest $\uf(n)\oplus \uf(m)$. This is in line with the observations of \cite{Schnabl:2008wj,
Hosomichi:2008jb}.

As an example let us consider the particular choice where $Z^A$ are $2\times 2$ complex matrices. A basis of
such matrices is provided by
\begin{equation}
T^a =
\left\{-\frac{i}{\sqrt{2}}\sigma_1,-\frac{i}{\sqrt{2}}\sigma_2,-\frac{i}{\sqrt{2}}\sigma_3,\frac{1}{\sqrt{2}}\one_{2
\times 2}\right\}\ ,
\end{equation}
where $a=1,2,3,4$, $\sigma_i$ are the Hermitian Pauli matrices: $\sigma_i\sigma_j = \delta_{ij}
+i\epsilon_{ijk}\sigma^k$ and the factor of $i$ is chosen to ensure that the structure constants $f^{abcd}$ are
real. In particular, using (\ref{ABJM3algebra}) and (\ref{metric}), one sees that
\begin{equation}
f^{abcd} = \frac{\pi}{k}\epsilon^{abcd} \qquad\textrm{and}\qquad {\rm Tr}(T^aT^b) = \delta^{ab}\ .
\end{equation}
Note that in this case $f^{abcd}$ is real and totally antisymmetric. This means that the Lagrangian ${\cal
L}_{\su(2)\times \su(2)}$  in fact has ${\cal N}=8$ supersymmetry and
 $\SO(8)$ R-symmetry and is precisely the Lagrangian  of
\cite{Bagger:2007jr}.

\subsection*{ From 3-algebras to the ABJM theory}

To obtain the $\U(n)\times \U(n)$ ABJM models that describe multiple M2-branes from the above we must gauge the rigid $\U(1)_B$ symmetry $Z^A \to e^{i\theta}Z^A$, $\psi_A\to e^{i\theta}\psi_A$ enjoyed by (\ref{niceaction}). Given any rigid supersymmetric theory with a global symmetry it is always possible to gauge this symmetry and preserve supersymmetry, provided that the supersymmetries commute with the global symmetries (otherwise the supersymmetries would have to become local and hence one would have to include gravity).

To gauge the $\U(1)_B$ we simply introduce an Abelian gauge field $B_\mu$ and redefine the covariant derivative $D_\mu$ to be
\begin{equation}\label{covder}
  D_\mu Z^A_a = \partial_\mu Z^A_a - \tilde A_{\mu}^b{}_aZ^A_b - i B_\mu \delta_a^b Z^A_b
\end{equation}
and similarly for $D_\mu \psi_{Aa}$ (${\bar Z}_A$ and $\psi^A$ have the opposite $\U(1)_B$ charge and hence the
sign of $\tilde A_\mu$ is flipped in $D_\mu {\bar Z}_A$ and $D_\mu \psi^A$). Under the $\U(1)_B$ gauge
transformation we have
\begin{equation}
B_\mu \to B_\mu +  \partial_\mu\theta
\end{equation}
and clearly the action is now invariant under $\U(1)_B$ gauge transformations so that the full gauge algebra is
$\su(n)\times \su(n)\times \uf(1)_Q\times \uf(1)_B$ (although again the $\U(1)_Q$ symmetry is trivial).

Our next step is to make the above action invariant under ${\cal N}=6$ supersymmetry. The transformations of $Z^A$, $\psi_A$ and $\tilde A_\mu^a{}_b$ remain the same, except that the covariant derivative now includes the $B_\mu$ gauge field. We will need $\delta B_\mu$ which we simply take to be
\begin{equation}
\delta B_\mu = 0\ .
\end{equation}
Since locally the theory is the same, the variation of the action is unchanged with the exception of terms in
the supervariation of the Fermion kinetic term involving $[D_\mu,D_\nu]$, which now includes a contribution
from $G_{\mu\nu} =
\partial_\mu B_\nu - \partial_\nu B_\mu$. Indeed we find
\begin{eqnarray}
\nonumber \delta{\cal L}_{\su(n)\times \su(n)}^{\textrm gauged} &=&-\frac{1}{2}
G_{\mu\nu}\bar\epsilon_{AB}\gamma^{\mu\nu}\psi^{Aa}Z^B_a+\frac{1}{2}
G_{\mu\nu}\bar\epsilon^{AB}\gamma^{\mu\nu}\psi_{Aa}\bZ_B^a\\
&=& -\frac{1}{2}\varepsilon^{\mu\nu\lambda}G_{\mu\nu}\bar\epsilon_{AB}\gamma_{\lambda}\psi^{Aa}Z^B_a
+\frac{1}{2}\varepsilon^{\mu\nu\lambda}\bar\epsilon^{AB} G_{\mu\nu}\bar\epsilon\gamma_\lambda\psi_{Aa}\bZ_B^a\
,
\end{eqnarray}
where we have used
$\gamma^{\mu\nu}=\varepsilon^{\mu\nu\lambda}\gamma_\lambda$. To
cancel this we introduce a new field $Q_\mu$ and a new term in the
Lagrangian
\begin{equation}\label{L}
 {\cal L}_{\uf(n)\oplus \uf(n)} =  {\cal L}_{\su(n)\oplus\su(n)}^{\textrm gauged} +
 \frac{k'}{8\pi}
 \epsilon^{\mu\nu\lambda}G_{\mu\nu}Q_\lambda\;,
\end{equation}
where in the first term on the right hand side we have included the $B_\mu$ gauge field and $k'$ is an as of
yet undetermined real constant. We see that this will be supersymmetric if we take
\begin{equation}
  \delta Q_\lambda =
  \frac{4\pi}{k'}\bar\epsilon_{AB}\gamma_{\lambda}\psi^{Aa}Z^B_a-\frac{4\pi}{k'}\bar\epsilon^{AB} \gamma_\lambda\psi_{Aa}\bar
  Z_B^a\ .
\end{equation}
The form for the supersymmetry transformations seems odd since $\delta B_\mu=0$ and hence
$[\delta_1,\delta_2]B_\mu=0$ so one might worry about closure. However on-shell we have $G_{\mu\nu}=0$ so
that, on-shell,
\begin{equation}
[\delta_1,\delta_2]B_\mu = v^\nu G_{\nu\mu}\qquad v^\nu=
\frac{i}{2}(\bar\epsilon_2^{CD}\gamma^\nu\epsilon^1_{CD})\ ,
\end{equation}
which is a translation and a $\U(1)_B$  gauge transformation. We must also check the closure on $Q_\mu$. Here we
find that
\begin{equation}
[\delta_1,\delta_2]Q_\mu = \frac{k'}{4\pi}v^\nu \varepsilon_{\mu\nu\lambda}(iZ^A_a D^\lambda \bZ^a_A -i
D^\lambda Z^A_a \bZ_A^a -\bar\psi^A_a\gamma^\lambda \psi_A^a) + D_\mu\Lambda\ ,
\end{equation}
where $\Lambda = \frac{k'}{4\pi} (\bar\epsilon_2^{AC}\epsilon_{1BC}-\bar\epsilon_1^{AC}\epsilon_{2BC} )\bar
Z^a_BZ^B_a$. Using the on-shell condition that comes from the Lagrangian
\begin{equation}
H_{\mu\nu} = -\frac{k'}{4\pi}\varepsilon_{\mu\nu\lambda}(iZ^A_a D^\lambda \bZ^a_A -i D^\lambda Z^A_a \bZ_A^a
-\bar\psi^A_a\gamma^\lambda \psi_A^a)\ ,
\end{equation}
where $H_{\mu\nu} =\partial_\mu Q_\nu -
\partial_\nu Q_\mu$,  we again find a translation with $\uf(1)_Q\times \uf(1)_B$ gauge
transformation
\begin{equation}
[\delta_1,\delta_2]Q_\mu = v^\nu H_{\nu \mu} + D_\mu\Lambda\ .
\end{equation}
Thus we see that $Q_\mu$, which started off life as a Lagrange multiplier for the constraint $G_{\mu\nu}=0$,
naturally inherits a $\uf(1)$ gauge symmetry of its own. The closure on the other fields remains unchanged from
the $\su(n)\times \su(n)$ Lagrangian, except that the connection now involves the $\uf(1)_B$ gauge field.

If we write $B_\mu = A^L_\mu-A^R_\mu$ and $Q_\mu= A^L_\mu+A^R_\mu$ then, up to a total derivative, the new term we have added is
\begin{equation}
{\cal L}_{\uf(1)\oplus\uf(1)\ CS} = \frac{k'}{4\pi} \epsilon^{\mu\nu\lambda}A^L_\mu\partial_\nu A^L_\lambda
-\frac{k'}{4\pi}\epsilon^{\mu\nu\lambda}A^R_\mu\partial_\nu A^R_\lambda\;,
\end{equation}
which is just the Chern-Simons Lagrangian for a $\uf(1)\oplus\uf(1)$ gauge theory.

We have therefore constructed a family of ${\cal N}=6$ Chern-Simons-matter Lagrangians with gauge fields that
take values in a $\uf(1)\oplus \su(n)\oplus \uf(1)\oplus \su(n)$ Lie-algebra and are parametrised by $k$ and
$k'$. From the point of view of supersymmetry the levels $k$ and $k'$ are arbitrary and although $k$ must be an
integer in the quantum theory, $k'$ need not be (indeed $k'$ can be absorbed into the definition of
$Q_\lambda$), \eg\;see \cite{Dunne:1998qy}. The possibility of choosing different levels was also pointed out
in \cite{Aharony:2008ug}.

With the choice\footnote{Here we agree with the literature \cite{Terashima:2008sy,Bandres:2008ry, Hosomichi:2008jb}   but normalise the $\U(n)$ generators with $T^a\in \SU(n)$ for $a = 1,...,N^2-1$ and $T^0 =   \one_{N \times N}$, such that the coefficients in the expression for the covariant derivative   (\ref{covder}) remain unchanged.}
\begin{equation}
k' = n k\ ,
\end{equation}
we see that the addition of the $\U(1)\times \U(1)$ Chern-Simons term simply converts the $\su(n)\times \su(n)$
level $(k,-k)$ Chern-Simons term ${\cal L}_{CS}$ with connection $\tilde A^a{}_b$ in the original Lagrangian
(\ref{CSterm}) into a $\uf(n)\times \uf(n)$ level $(k,-k)$ Chern-Simons term with connection $\tilde
A_\mu^{L/R}+ iA^{L/R}_\mu$. In terms of $A^{R/L}_\mu$, we have
\begin{equation}
\delta A^R_\lambda =\delta A^L_\lambda=
\frac{2\pi}{nk}\bar\epsilon_{AB}\gamma_{\lambda}\psi^{Aa}Z^B_a-\frac{2\pi}{nk}\bar\epsilon^{AB} \gamma_\lambda\psi_{Aa}\bar
Z_B^a\ .
\end{equation}
Taking the global gauge group to be $\U(n)\times \U(n)$ we have constructed the ${\cal N}=6$ ABJM theory
\cite{Aharony:2008ug}.

Finally we mention a crucial subtlety: the decomposition of $\U(n)$ is not strictly in terms of $\SU(n)\times
\U(1)$. In particular given any pair $\omega\in \U(1)$ and $A_0\in \SU(n)$ we obtain an element $A = \omega A_0
\in \U(n)$. However the inverse map is not unique since, for a given $A\in \U(n)$, we have
\begin{equation}
\omega^n = \det (A)\ ,\qquad A_0 = \omega^{-1} A\ ,
\end{equation}
and hence there are $n$ solutions for $\omega$ and $A_0$ related by $\omega\to e^{2\pi   i/n}\omega$, $A_0\to e^{-2\pi i/n} A_0$. Thus the map from $ \U(1)\times \SU(n)\to \U(n)$ is an $n$-fold cover and so the isomorphism is
\begin{equation}
\U(n)\simeq \frac{\SU(n)\times \U(1)}{\mathbb{Z}_n} \;.
\end{equation}
Although these modifications do not change anything at the level of the Lagrangian or the classical theory,
they do change the quantisation conditions for the various fluxes, as we shall see in the next section, which
will be important in the next section when we calculate the moduli space of the theory in order to compare with
the answer expected from M-theory.

\subsection*{Dual Photon Formulation}
Having arrived at the standard form for the ABJM theory we can take a step back and consider the equivalent
Lagrangian (\ref{L}), but once again with $k'=nk$. Integrating by parts and discarding a boundary term leads to
\begin{equation}
 {\cal L}_{\uf(n)\times \uf(n)} =  {\cal L}_{\su(n)\times su(n)}^{\textrm gauged} +
\frac{nk}{4\pi}
 \varepsilon^{\mu\nu\lambda}B_\mu\partial_\nu Q_\lambda\ .
\end{equation}
Next we introduce a Lagrange multiplier term\footnote{Aspects of this procedure have also appeared   in \cite{Gustavsson:2009pm,Kwon:2009ar}.}
\begin{equation}
 {\cal L}_{\uf(n)\times \uf(n)}=  {\cal L}_{\su(n)\times \su(n)}^{\textrm gauged} +
 \frac{nk}{8\pi}
 \varepsilon^{\mu\nu\lambda}B_\mu H_{\nu\lambda} +\frac{n}{8\pi} \sigma\varepsilon^{\mu\nu\lambda} \partial_{\mu}H_{\nu\lambda}\ .
\end{equation}
Integrating the last term by parts we find
\begin{equation}
 {\cal L}_{\uf(n)\oplus \uf(n)} =  {\cal L}_{\su(n)\oplus \su(n)}^{\textrm gauged} +
 \frac{nk}{8\pi}
 \varepsilon^{\mu\nu\lambda}B_\mu H_{\nu\lambda}
 - \frac{n}{8\pi}\varepsilon^{\mu\nu\lambda}\partial_\mu\sigma H_{\nu\lambda}\ .
\end{equation}
We can now integrate out $H_{\mu\nu}$ to see that
\begin{equation}
B_\mu = \frac{1}{k}\partial_\mu \sigma\ .
\end{equation}
Thus under a $\U(1)_B$ gauge transformation we find
\begin{equation}\label{fix}
\sigma \to \sigma +  {k}\theta\ .
\end{equation}
Substituting back  we find that the $\uf(n)\oplus \uf(n)$ Lagrangian is equivalent to the ${\su(n)\oplus
\su(n)}$ Lagrangian with new variables:
\begin{equation}
{\cal L}_{\uf(n)\oplus \uf(n)}(Z^A,\psi_A,\tilde A_\mu^a{}_b,B_\mu,Q_\mu) \cong {\cal
L}_{\su(n)\oplus\su(n)}(e^{\frac{i}{k}\sigma}Z^A,e^{\frac{i}{k}\sigma}\psi_A,\tilde A_\mu^a{}_b)\;.
\end{equation}
In particular the variables $\hat Z^A=e^{\frac{i}{k}\sigma}Z^A$ and $\hat \psi_A=e^{\frac{i}{k}\sigma}\psi_A$ are $\U(1)\times \U(1)$ gauge invariant.

Finally, we need to determine the periodicity of $\sigma$ which follows from a quantisation condition on the flux $H$. Let us review the familiar Dirac quantisation rule. We start by considering the phase induced by the parallel transport over a closed path $\gamma$ of a field, $\Psi$, that couples to a $\U(1)$ field $A_\mu$ through $D_\mu\Psi =\partial_\mu \Psi - iA_\mu \Psi$. We find that the resulting wavefunction is related to the initial wavefunction by a $\U(1)$ transformation
\begin{equation}
\Psi_\gamma = e^{i\oint_\gamma A}\Psi_0=e^{i\int_D F}\Psi_0\;,
\end{equation}
where $D$ is a two-dimensional surface whose boundary is $\gamma$. However the choice of $D$ is not unique.  Given any two such choices $D$ and $D'$ we require that the phase, viewed as an element of the gauge group $\U(1)$, is the same. This implies that
\begin{equation}
e^{i\int_{D-D'} F}=1
\end{equation}
and hence $\int_\Sigma F \in 2\pi\mathbb{Z} $, where $\Sigma=D-D'$ is any closed surface. However in our case the gauge group is $(\U(1)\times \mathrm{SU}(n))/\mathbb{Z}_n$ and we need only require that $ \int_\Sigma F \in \frac{2\pi}{n}\mathbb{Z}$, {\it i.e.} the $\U(1)$ phases computed by two different paths must be equal modulo $\mathbb{Z}_n$.  Thus we see that the quantisation condition is
\begin{equation}\label{quantcon1}
\int  dF_{L/R} \in  \frac{2\pi}{n} \mathbb{Z}\ .
\end{equation}
This fractional flux quantization condition arises because the global gauge group is $(\SU(n)\times \SU(n))/\mathbb Z_n$ instead of $\SU(n)\times \SU(n)$, with $\mathbb{Z}_n$ the relative centre of the two $\SU(n)$ factors. Thus we refer to the resulting Chern-Simons matter theory as the $(\SU(n)\times \SU(n))/\mathbb Z_n$-theory.\footnote{For theories with bifundamental   matter the $(\SU(n)\times \SU(n))/\mathbb Z_n$ group, where the centre of one $\SU(n)$ factor is   identified with the inverse centre of the other, is indistinguishable from $\SU(n))/\mathbb Z_n   \times \SU(n))/\mathbb Z_n \simeq \mathrm{PSU}(n) \times\mathrm{PSU}(n)$.}  This should be compared with a theory with the same ${\cal L}_{\su(n)\oplus\su(n)}$ Lagrangian but global $\SU(n)\times \SU(n)$ gauge symmetry and no fractional flux quantisation which we refer to as the $\SU(n)\times \SU(n)$-theory.

After integrating out $H$, we are left with the condition $B = \frac{1}{k}d\sigma$. Therefore, locally,
$F_L-F_R=dB$ vanishes so that $F_L$ and $F_R$ must have the same flux. Note that we do not require that
$\sigma$ is globally defined so there can be a non-zero Wilson line for the gauge field $B$. However, since
$F_L-F_R=dB=0$ in any open set where $\sigma$ is single-valued, it follows that $F_L=F_R$ globally. This
generalises the flux quantisation argument of \cite{Martelli:2008si} to allow for a nonvanishing but trivial
gauge field and applies to the full theory, not just the moduli space. Since $H = F_L+F_R$ we have
\begin{equation}\label{Hflux}
\int d H=\int \frac{1}{2}\epsilon^{\mu\nu\lambda}\partial_\mu H_{\nu\lambda} \in \frac{4\pi } {n}\mathbb{Z}
\end{equation}
and $\sigma$ has period $2\pi$. Note that since $e^{i\theta}$ is a $\U(n)$ transformation, $\theta$ also has period $2\pi$. Thus we can fix the $\U(1)_B$ symmetry using (\ref{fix}) and set $\sigma=0\ {\rm mod }\ 2\pi$. However, this periodicity imposes an additional identification on the $\U(1)$-invariant fields
\begin{equation}\label{U(1)b}
\hat Z^A \cong e^{\frac{2\pi  i}{k}}\hat Z^A\qquad \textrm{and} \qquad \hat \psi_A \cong
e^{\frac{2\pi  i}{k}}\hat \psi_A\ .
\end{equation}
We are therefore told that the $\U(n)\times \U(n)$ ABJM theory is equivalent to a $\mathbb{Z}_k$ identification on the $(\SU(n)\times \SU(n))/\mathbb Z_n$-theory. Note that the $\mathbb Z_n$ quotient arises here as the relative part of the two $\mathbb Z_n$ factors from $\U(n)\simeq (\U(1)\times \SU(n))/\mathbb Z_n$.

However we should be careful: Our discussion so far has been largely based on local aspects of the theory and since $\U(n)$ is not globally the same as $\U(1)\times \SU(n)$ there could be obstructions at a global level.  We will see in the following that the $\U(n)\times \U(n)$ theories can only be viewed as $\mathbb{Z}_k$ identifications when $n$ and $k$ are coprime. In particular, for $k=1$ the $\mathbb{Z}_k$ identification is clearly trivial and one simply has the $(\SU(n)\times \SU(n))/\mathbb Z_n$-theory.

Note that, had we considered instead a $\U(1) \times \SU(n)\times \U(1)\times \SU(n)$ gauge theory, we would
not have been able to use the fractional flux quantisation condition and $\sigma$ would have had period
$2\pi/n$. In addition, we would have been free to have any integer value for the $\U(1)$ level $k'$ and as a
result we would find a $\mathbb{Z}_{k'}$ identification. From this perspective we would arrive at a
$\SU(n)\times \SU(n)$-theory by starting with $\U(1)\times \SU(n)\times \U(1)\times \SU(n)$ but take $k'=k$ and
the usual Dirac quantisation. However, as we will see in the next section, the moduli space of the resulting
theory would then not be the same as the $\U(n)\times \U(n)$ ABJM models due to the different flux quantisation
condition on the $\SU(n)$ factor. Finally, one might consider other quantisation conditions which lead to
different moduli spaces \cite{Berenstein:2009sa}.

\subsection*{Moduli Space of $n=2$ theories}

To test the above analysis it is insightful to compute the moduli space of the $(\SU(n)\times \SU(n))/\mathbb
Z_n$-theory and then compare with the $\U(n)\times \U(n)$ answer. To begin with, we consider the $n=2$ case in
detail.

We observe that the solutions to $V=0$ are obtained by taking $[Z^A,Z^B;\bar Z_C]=0$ for all $A,B,C$. This is
solved by taking the $Z^A$, which are $2\times 2$ matrices, to be mutually commuting. Recall that the $Z^A$ are
in the bi-fundamental representation so that under a gauge transformation
\begin{equation}
Z^A \cong g_LZ^Ag_R^{-1}\;.
\end{equation}
Thus, modulo gauge transformations, we can take without loss of generality
\begin{equation}\label{vac}
Z^A  = \frac{1}{\sqrt 2}r_1^A - \frac{i}{\sqrt 2}r_2^A\sigma_3\ .
\end{equation}
The gauge symmetries that preserve this form, for generic $r_1^A,r_2^A$, must satisfy
\begin{equation}
g_Lg_R^{-1} = a +ib\sigma_3\qquad g_Li\sigma_3g_R^{-1} = c +id\sigma_3
\end{equation}
for arbitrary constants $a,b,c,d$.  The first condition can be used to deduce that
\begin{equation}
g_L =e^{i\theta\sigma_3} g_R\ ,
\end{equation}
for an arbitrary $\theta$, whereas the second condition puts a constraint on $g_R$
\begin{equation}
g_Ri\sigma_3 g_R^{-1}= e^{i\theta'\sigma_3} ,
\end{equation}
for an arbitrary $\theta'$. Since the left hand side is traceless we see that this is only possible if $\theta'=\pm \pi/2$ so that
\begin{equation}\label{gaugeconst}
g_Ri\sigma_3 g_R^{-1}= \pm i\sigma_3 .
\end{equation}
Thus  $g_R$  is generating a discrete identification
\begin{equation}
r_2^A\cong -r_2^A
\end{equation}
and one should think of $r_1^A$ as the centre-of-mass coordinate, while $r_2^A$ as the relative separation between two indistinguishable M2-branes. To this end we write
\begin{equation}\label{r1r2}
r_1^A = \frac{1}{2}(z_1^A + z_2^A)\qquad\mathrm{and}\qquad r_2^A= \frac{i}{2}(z_1^A-z_2^A)\;,
\end{equation}
so that the $g_R$ transformation is now $z_1^A \leftrightarrow z_2^A$. In addition we have a continuous $\U(1)$ action generated by $g_L=e^{i\theta\sigma_3}$. This acts on $z_1^A$ and $z^A_2$ as
\begin{equation}
z_1^A\to e^{i \theta} z_1^A\;,\qquad z_2^A\to e^{-i\theta} z_2^A\ .
\end{equation}
The subtle part of the calculation comes from considering the
continuous gauge symmetries $g=e^{i\theta}$.  Reducing to the
moduli space fields with with $A_\mu = \tilde A^3_{L\mu} \sigma^3$ and $\tilde A_\mu = \tilde A^3_{R\mu} \sigma^3$, we find that the Chern-Simons action  (\ref{CSterm}) becomes
\begin{equation}
{\cal L} = -{\cal D}_\mu z_1^A{\cal D}^\mu \bar z_{1A}-{\cal D}_\mu
z_2^A{\cal D}^\mu \bar z_{2A}+
\frac{k}{2\pi}\epsilon^{\mu\nu\lambda}\tilde
A^3_{L\mu}\partial_\nu\tilde
A^3_{L\lambda}-\frac{k}{2\pi}\epsilon^{\mu\nu\lambda}\tilde
A^3_{R\mu}\partial_\nu\tilde A^3_{R\lambda}\;,
\end{equation}
where
\begin{equation}
{\cal D_\mu} z_1^A = \partial_\mu z_1^A - i(\tilde
A_{L\mu}^{3}-\tilde A_{R\mu}^{3})z_1^A \ ,\qquad {\cal D_\mu} z_2^A
=
\partial_\mu z_2^A + i(\tilde A_{L\mu}^{3}-\tilde
A_{R\mu}^{3})z_2^A\;.
\end{equation}
Following the previous discussion we write $\tilde B_\mu = \tilde A_{L\mu}^{3}-\tilde A_{R\mu}^{3}$ and $\tilde Q_\mu = \tilde A_{L\mu}^{3}+\tilde A_{R\mu}^{3}$ so that the moduli space Lagrangian is
\begin{equation}
{\cal L} = -{\cal D}_\mu z_1^A{\cal D}^\mu \bar z_{1A}-{\cal D}_\mu
z_2^A{\cal D}^\mu \bar z_{2A} +
\frac{2k}{8\pi}\epsilon^{\mu\nu\lambda}\tilde B_\mu \tilde
H_{\nu\lambda}\;,
\end{equation}
where now $\tilde H_{\nu\lambda} = \partial_\nu\tilde Q_\lambda-\partial_\lambda \tilde Q_\nu$. We can introduce a Lagrange multiplier term
\begin{equation}
{\cal L} =-{\cal D}_\mu z_1^A{\cal D}^\mu \bar z_{1A}-{\cal D}_\mu z_2^A{\cal D}^\mu \bar z_{2A}+
\frac{2k}{8\pi}\epsilon^{\mu\nu\lambda}\tilde B_\mu \tilde H_{\nu\lambda} +
\frac{2}{8\pi}\chi\epsilon^{\mu\nu\lambda}
\partial_\mu \tilde H_{\nu\lambda}\ .
\end{equation}
Integrating out $\tilde H_{\mu\nu}$ gives $\tilde B_\mu =\frac{1}{k}\partial_\mu\chi$ and the Lagrangian can be written as
\begin{equation}
{\cal L} = -{\partial}_\mu \tilde z_1^A {\partial }^\mu \bar {\tilde
z}_{1 A}-{\partial}_\mu \tilde z_1^A {\partial }^\mu \bar {\tilde
z}_{2 A}\;,
\end{equation}
where $\tilde z_1^A = e^{i\chi/k}z_1^A$ and $\tilde z_2^A = e^{-i\chi/k}z_2^A$ are gauge invariant.

It is once again necessary to determine the periodicity of the dual photon $\chi$. The argument here is identical to what was discussed around Eq.\;(\ref{Hflux}). Namely, since $d\tilde B=0$ we have that $\tilde F^3_L=\tilde F^3_R$, where $\tilde F^3_{L/R} = d\tilde A_{L/R}^3$, and the quantisation condition is
\begin{equation}\label{quantcon2}
\int d\tilde F^3_{L/R} \in \frac{2\pi }{2}\mathbb{Z}\;.
\end{equation}
The factor of $2$ in the denominator arises because the gauge group is $(\mathrm{SU}(2)\times \U(1))/\mathbb{Z}_2$, in the same manner that $n$ appeared in (\ref{quantcon1}). Thus, since $\tilde H =\tilde F^3_L+\tilde F^3_R$ we have that
\begin{equation}
\int d\tilde H = \int \frac{1}{2}\epsilon^{\mu\nu\lambda}\partial_\mu\tilde H_{\nu\lambda} \in
\frac{4\pi}{2}\mathbb{Z}
\end{equation}
and hence $\chi$ has period of $2\pi$. We conclude that the vacuum moduli space scalars are subject to the identification
\begin{equation}
z^A_1 \cong e^{\frac{2\pi i}{k}}z_1^A\;,\qquad z^A_2 \cong e^{-\frac{2\pi i}{k}}z_2^A\ .
\end{equation}

In summary, we find that the sum of identifications on the vacuum moduli space, including the ones coming from (\ref{U(1)b}), act as
\begin{eqnarray}\label{identifications}
\nonumber  g_{\U(1)}:\quad z^A_1 &\cong& e^{\frac{2\pi i}{k}}z_1^A\;, \qquad z^A_2
\cong e^{\frac{2\pi i}{k}}z_2^A\\
 g_{12}:\quad z^A_1 &\cong& z^A_2 \\
\nonumber g_{\SU(2)}:\quad z^A_1 &\cong& e^{\frac{2\pi i}{k}}z_1^A\;, \qquad z^A_2 \cong e^{-\frac{2\pi
i}{k}}z_2^A\;.
\end{eqnarray}
The first one is a $\mathbb{Z}_k$ coming from integrating out the $\U(1)_B$, and acts on the whole theory, not just the moduli space. The other two are consequences of the $(\SU(n)\times \SU(n))/\mathbb Z_n$ gauge symmetry acting on the vacuum moduli space and generate the dihedral group of order $2k$, $\mathbb{D}_{k}\simeq \mathbb{Z}_2 \ltimes \mathbb{ Z}_k$. This is consistent with the calculation in \cite{Lambert:2008et,Distler:2008mk} which found $\mathbb{D}_{2k}$, since the difference $k\to 2k$ arises because a fractional quantisation condition was not allowed, corresponding to an $\SU(2)\times\SU(2)$ global gauge group.

We now need to compare these moduli space identifications with the answer for the $\U(2)\times \U(2)$ ABJM
theory that describes two indistinguishable M2-branes in $\mathbb{R}^8/\mathbb{Z}_k$, that is
\begin{equation}
{\cal M}_k = \frac{(\mathbb{R}^8/\mathbb{Z}_k)\times (\mathbb{R}^8/\mathbb{Z}_k)}{\mathbb{Z}_2}\ .
\end{equation}
In this case the moduli space quotient group is generated by
\begin{eqnarray}
\nonumber  g_1:\quad z^A_1 &\cong& e^{\frac{2\pi i}{k}}z_1^A\;, \qquad z^A_2
\cong z_2^A\\
 g_{12}:\quad z^A_1 &\cong& z^A_2 \\
\nonumber g_{2}:\quad z^A_1 &\cong& z_1^A\;, \qquad z^A_2 \cong e^{\frac{2\pi i}{k}}z_2^A\;.
\end{eqnarray}
Here we see that $g_{\U(1)}=g_1g_2$ and $g_{\SU(2)}=g_2^{-1}g_{1}$. However in order to invert these relations we need to
solve $g^2_1=g_{\U(1)}g_{\SU(2)}$ and $g^2_2= g_{\U(1)} g_{\SU(2)}^{-1} $, {\it i.e.} take the square root in the
group generated by $g_{\U(1)}$ and $g_{\SU(2)}$. A short calculation shows that this is only possible if $k$ is
odd. Thus we conclude the we obtain the correct moduli space only when $k$ is odd.

The value $k=1$ is special: The orbifold action is trivial and the moduli space of the $\mathrm{SU}(2)\times
\mathrm{SU}(2)$-theory is the one for 2 M2-branes in flat space. As a by-product we see that for $k=1$ the
$\mathcal N=6$ $\U(2)\times \U(2)$-theory in fact has $\mathcal N = 8$ supersymmetry. This has also been shown
with the help of monopole operators in \cite{Aharony:2008ug,Gustavsson:2009pm,Kwon:2009ar}, although here the
physics also have a formulation in terms of the manifestly $\mathcal{N} = 8$ supersymmetric, local Lagrangian.

\subsection*{Moduli Space In General}

We will now see that the problem we faced for $n=2$ and $k$-even extends more generally. For a general $n$ the
vacuum moduli space is obtained by setting
\begin{equation}
Z^A = {\rm diag}(z_1^A,...,z_n^A)\ .
\end{equation}
If we consider gauge transformations of the form $g_L=g_R$ then $Z^A$ behaves as if it were in the adjoint of
$\SU(n)$ and hence cannot tell the difference between the $\SU(n)$ and $\U(n)$ theories. The result is that the
gauge transformations which preserve the form of $Z^A$ simply interchange the eigenvalues $z_i^A$ leading to
the symmetric group acting on the $n$ M2-branes.

Next we can consider transformations in the diagonal subgroup of $\SU(n)$ or $\U(n)$. These act to rotate the
phases of the $z^A_i$, however in the $\SU(n)$-theory they only do so up to the constraint that the diagonal
elements must have unit determinant. In the $\U(n)$-theory this is not the case and there are $n$ independent
$\U(1)$'s, one for each $z_i^A$,  and each of these $\U(1)$'s leads to a $\mathbb{Z}_k$ identification on the
moduli space. Thus for $\U(n)$ we indeed see that we find $n$ commuting copies of $\mathbb{Z}_k$ along with the
symmetric group acting on the $z_i^A$.

For the $\SU(n)$-theory, even including the $\mathbb{Z}_k$ action of $\U(1)_B$, this will not always be the
case. In particular, note that since the determinant of the gauge transformations coming from $\SU(n)$ is
always one we have, for an arbitrary element of the moduli space orbifold group,
\begin{equation}
{\rm det} (g_{\U(1)}^{l_B} g_{0}) = {\rm det} (g_{\U(1)}^{l_B} ) = e^{2\pi i nl_B/k}\ .
\end{equation}
Here $g_{0}$ represents a generic element of the moduli space orbifold group obtained in the $(\SU(n)\times
\SU(n))/\mathbb Z_n$-theory. On the other hand, the moduli space orbifold group of the $\U(n)$-theory generated
by $n$ independent $\U(1)$'s has
\begin{equation}
{\rm det} (g_{1}^{l_1}...g_n^{l_n}) =  e^{2\pi i(l_1+...+l_n)/k}\ .
\end{equation}
If these two theories are to give the same moduli space then we must be able to have $e^{2\pi i
(l_1+...+l_n)/k}=e^{2\pi i n l_B/k}$ for any possible combination of $l_i$'s. Thus we are required to solve
\begin{equation}\label{llb}
l= n\;l_B\ {\rm mod}\;k\ ,
\end{equation}
for $l_B$ as a function of $l,n,k$, where $l = l_1+...+l_n$ is arbitrary. Hence, if this equation can be solved
for $l_B$ then $g_{0}=e^{-2\pi i l_B/k}g_{1}^{l_1}...g_n^{l_n}$ is an element of $\SU(n)$ and can arise from
the vacuum moduli space quotient group of the$(\SU(n)\times \SU(n))/\mathbb Z_n$-theory.

We will now show that (\ref{llb}) has solutions for all $l$ {\bf iff} $n$ and $k$ are coprime. In general
the solution is $l_B = (l-pk)/n$ for any $p\in \mathbb{Z}$; however we require that $l_B$ is an integer. It is
clear that we may view $l,k$ and $p$ as elements of $\mathbb{Z}/\mathbb{Z}_n$ and we are therefore required
to solve the following equation for $p$
\begin{equation}
l=pk \ {\rm mod}\ n\ .
\end{equation}
This always has solutions if the map $\varphi:p\mapsto pk$  is surjective on $\mathbb{Z}/\mathbb{Z}_n$. Since
$\mathbb{Z}/\mathbb{Z}_n$ is a finite set this will be the case {\bf iff} $\varphi$ is also injective. Thus we wish to
show that $pk=p'k\ {\rm mod}\ n$ implies $p=p'$. This is equivalent to showing that $qk=0\ {\rm mod}\ n$
implies $q=0\ {\rm mod}\ n$. Now suppose that $qk=rn$. If $k$ and $n$ are coprime then all the prime
factors of $k$ must be in $r$  and all the prime factors of $n$ must be in $q$. Thus $q=0\ {\rm mod}\ n$. On
the other hand if $k$ and $n$ have a common factor $d$ then we find a non-zero solution by taking $q =n/d$ and
$r=k/d$. Thus $qk=0\ {\rm mod}\ n$ has no non-trivial solutions for $q$ {\bf iff} $n$ and $k$ are coprime.

This result can been restated as follows: Although locally $\U(n)\simeq \U(1)\times \SU(n)$, this is not true
globally. Even though the Lagrangian is defined by local information at the Lie-algebra level, the  map we
constructed, reducing the $\U(n)\times \U(n)$-theory to a $\mathbb{Z}_k$ quotient of the $(\SU(n)\times
\SU(n))/\mathbb Z_n$-theory, involves finite gauge transformations and is therefore sensitive to global
properties of $\U(n)$. The above discussion shows that the vacuum moduli space quotient group of the
$\U(n)\times \U(n)$ theories is not of the form $\mathbb{Z}_k\times G_0$, where $G_0\subset\SU(n)$, unless $n$
and $k$ are relatively prime.

We have therefore shown that if $n$ and $k$ have a common factor then the vacuum moduli spaces for the two
theories do not agree, as there is a global obstruction to mapping the $\U(n)\times \U(n)$-theory to a
$\mathbb{Z}_k$ quotient of the $(\SU(n)\times \SU(n))/\mathbb Z_n$-theory. On the other hand, if $n$ and $k$
are coprime then the vacuum moduli space calculated in the $(\SU(n)\times \SU(n))/\mathbb Z_n$-theory, along
with the $\mathbb{Z}_k$ identification coming from $\U(1)_B$, agrees with the vacuum moduli space of the
$\U(n)\times \U(n)$-theory. It is therefore natural to conjecture that in these cases the $\U(n)\times \U(n)$
theories are $\mathbb{Z}_k$ quotients of the $(\SU(n)\times \SU(n))/\mathbb Z_n$ theories.

\subsection*{The moduli space of the $k=2$ $\SU(2)\times\SU(2)$-theory}

On a related note, the moduli space of $\SU(2)\times \SU(2)$ ${\cal N}=8$ theories was calculated in
\cite{Lambert:2008et,Distler:2008mk} and found to be $(\mathbb {R}^8\times \mathbb{R}^8)/ \mathbb{ D}_{2k}$,
where $\mathbb{D}_{2k} \simeq \mathbb{Z}_2 \ltimes \mathbb{Z}_{2k}$ the dihedral group of order $4k$. As we
have already mentioned, the extra factor of $2$ arises due to the standard Dirac quantisation condition when
the global gauge group is $\SU(2)\times\SU(2)$.

For the particular case of $k=2$ one has \cite{Lambert:2008et,Distler:2008mk}
\begin{eqnarray}
 g_{12}:\quad z^A_1 &\cong& z^A_2 \\
\nonumber g_{\SU(2)}:\quad z^A_1 &\cong& i z_1^A\;, \qquad z^A_2 \cong -i z_2^A\;.
\end{eqnarray}
Note that because of the modified flux quantisation condition this agrees with $k=4$ in (\ref{identifications}).
Interestingly, by reverting back to the $r^A_1, r^A_2$ variables of (\ref{r1r2}) we have
\begin{eqnarray}
 g_{12}:\quad r^A_1 \cong r^A_1 &,& r^A_2 \cong -r^A_2 \\
\nonumber g_{\SU(2)}:\quad r^A_1 \cong r^A_2 &,&  r^A_2 \cong - r_1^A\;.
\end{eqnarray}
Although these variables might look contrived from the perspective of the ABJM theory they arise very naturally
in the $\SO(4)$ formulation \cite{Lambert:2008et}. From these we can construct
\begin{eqnarray}
 g_{12}:\quad  r^A_1 \cong r^A_1 &,&r^A_2 \cong -r^A_2 \\
\nonumber g_{12}\; g^2_{\SU(2)}:\quad r^A_1 \cong -r^A_1 &,& r^A_2 \cong r^A_2 \\
 \nonumber  g_{\SU(2)}\; g_{12}:\quad r^A_1 \cong r^A_2 &,& r^A_2 \cong r^A_1\;.
\end{eqnarray}
These identifications are the ones expected for the moduli space $(\mathbb R^8/\mathbb Z_2 \times \mathbb
R^8/\mathbb Z_2)/\mathbb Z_2$ of 2 M2-branes on a $\mathbb Z_2$ orbifold singularity of M-theory, as also shown
in \cite{Lambert:2008et}.

For $k=2$ the $\SU(2)\times \SU(2)$-theory was interpreted in \cite{Lambert:2008et,Distler:2008mk} as the IR
limit of an $\SO(5)$ gauge theory describing two D2-branes on an $\widetilde{\mathrm{O}2}^+$ orientifold of
type IIA string theory, which is an M-theory $\mathbb{Z}_2$ orbifold with discrete torsion. However, there also
exists another type IIA orientifold denoted $\mathrm{O}2^-$ and, as was pointed out in \cite{Aharony:2008ug},
corresponding to an $\mathrm{O}(4)$-theory on the D2-brane worldvolume, which in the IR lifts to an M-theory
orbifold without torsion. This has an indistinguishable moduli space from the $\SO(5)$ case, since the extra
fractional brane in the latter is stuck at the fixed point and does not contribute to the moduli space
dynamics. The orbifolds with and without torsion are the only expected IR fixed points with ${\cal N}=8$
supersymmetry and $(\mathbb R^8/\mathbb Z_2 \times \mathbb R^8/\mathbb Z_2)/\mathbb Z_2$ moduli space and
correspond to the $\U(2)\times \U(2)$ ABJM and $\U(2)\times \U(3)$ ABJ theories respectively. Given the
similarity between the Lagrangians, manifest symmetries (such as Parity) and the agreement between the  moduli
space calculations, it is also natural to conjecture that the $n=2,k=2$ $(\SU(2) \times \SU(2))/\mathbb
Z_2$-theory is equivalent to the $k=2$ ABJM theory\footnote{Note that the ABJM theory at $n=2$, $k=2$ is not
related to the $\SU(2)\times \SU(2)$-theory as discussed in the previous section.} and therefore the IR fixed
point of the the maximally supersymmetric $\mathrm{O}(4)$ gauge theory in 2+1d.

\subsection*{Summary}

In this paper we have discussed the relation of $\U(n)\times \U(n)$ ABJM theories to $(\SU(n)\times \SU(n))/\mathbb Z_n$ theories. In particular we showed that locally, at the level of Lagrangians, the $\U(1)_B$ gauge symmetry could be integrated out to give an $(\SU(n)\times \SU(n))/\mathbb Z_n$-theory along with a $\mathbb{Z}_k$ identification on the fields. However we also saw that there was a global obstruction to this when $n$ and $k$ are not coprime.

As a result we found that the $\U(2)\times \U(2)$ ABJM theories can be viewed as $\mathbb{Z}_k$ quotients of the $(\SU(2) \times \SU(2))/\mathbb Z_2$, ${\cal N}=8$ theories when $k$ is odd. In particular for $k=1$ they are identical. However if one considers the $\SU(2)\times \SU(2)$, ${\cal   N}=8$-theory of \cite{Bagger:2007jr,Gustavsson:2007vu} at $k=2$, then this has the same moduli space, global supersymmetries and manifest Parity as the $k=2$ $\U(2)\times \U(2)$ ABJM theory. Thus we conjectured that these two theories are equivalent and the original ${\cal   N}=8$, $\su(2)\times \su(2)$ Lagrangian of \cite{Bagger:2007jr,Gustavsson:2007vu} can be used to define quantum theories for two M2-branes on $\mathbb{R}^8$ and $\mathbb{R}^8/\mathbb{Z}_2$ (without discrete torsion) when $k=1$ or $k=2$ respectively and with all the symmetries manifest.

\subsection*{Acknowledgements}

We would like to thank Ofer Aharony, Jon Bagger, Sunil Mukhi and David Tong for useful discussions and comments, as well as the Fundamental Physics UK 3.0 for providing a stimulating environment towards the completion of this work. The authors are supported by the STFC grant ST/G000395/1.

\bibliographystyle{utphys}
\bibliography{ABJMvsBL}

\end{document}